==================

\documentstyle [12pt]{article}
\title{
\hspace{3.8truein}{\small UBCTP-92-34}\\
{\bf SOLVING
THE CONFORMAL BOOTSTRAP: FOUR-FERMION
 INTERACTIONS
 AT $2<d<4$}}
\author{Wei Chen$^\dagger$, Yuri Makeenko$^\ddagger$ and
Gordon W.\ Semenoff$^\dagger$\\
\\
$^\dagger$ Department of Physics, University of British Columbia\\
Vancouver, British Columbia, Canada V6T 1Z1\\
\\
$^\ddagger$ Institute of Theoretical and Experimental Physics\\ 117259
Moscow, Russia}

\vspace{0.2truein}
\date{{\it PACS No.: 11.10.Gh}}

\begin{document}
\maketitle

\vspace{0.1truein}
\begin{abstract}
We formulate the conformal bootstrap approach to four--fermion theory
at its strong coupling fixed point in dimensions $2<d<4$.  We present
a solution of the bootstrap equations in the five--vertex
approximation.  We show that the bootstrap approach gives a
particularly simple way to obtain next to leading order corrections to
critical exponents in the large $N$ expansion and present the values
of the anomalous dimensions of the fermion field $\psi$ and the
composite $\bar\psi\psi$ to order $1/N^2$.
\end{abstract}

\baselineskip=20.0truept
\newpage

Conformal symmetry describes the critical behavior of quantum field
theories or statistical systems at points of second order phase
transition.  It is widely used in two dimensions where it provides
exact solutions to a variety of non--trivial interacting field
theories.  In this Letter we shall demonstrate that it can also be
useful in higher dimensions.  If a field theory has a fixed point of
the renormalization group flow, the dynamics at that point is
conformally invariant.  In any spacetime dimensions, conformal
symmetry determines the form of the two and three-point Green
functions of the field theory up to a few constants, the scaling
dimension of the field operators and the value of the coupling
constant at the fixed point.  This information, when combined with
Dyson--Schwinger equations can provide a powerful tool for the
analysis of field theories.

In fact, in a field theory with only three-point coupling constants,
the Dyson--Schwinger equations which relate two and three point
functions could in principle be used to determine all anomalous
dimensions in terms of the coupling constants, which are then left to
be determined by higher order equations.  Unfortunately, this direct
approach is plagued by technical difficulties involving indeterminate
expressions.  This problem was partially solved by Parisi
\cite{parisi} who showed how to use the information of the lower order
Dyson--Schwinger equations to derive bootstrap equations for the
effective coupling constants.  The resulting equations can not be
solved exactly, but can be used to find interesting approximate
solutions of a field theory.

In this Letter, we shall consider the example of a fermionic field
theory with a four-fermion interaction, $-\frac{\lambda^2}{2}
\left(\sum_{i=1}^N\bar\psi_i\psi_i\right)^2$ in spacetime
dimensions $2<d<4$.  By introducing an auxillary scalar field $\phi=
\lambda\sum_{i=1}^N\bar\psi_i\psi_i$ (with no kinetic term), this is equivalent
to a model with a Yukawa vertex
$-\lambda\phi\sum_{i=1}^N\bar\psi_i\psi_i+\phi^2/2$. The
renormalization group flow in this model has an infrared stable fixed
point at $\lambda=0$ and there is a long-standing conjecture that it
also has an ultraviolet stable fixed point at some $\lambda_*>0$.
This fixed point is seen both in the large $N$ and in the $\epsilon$
expansion.  Also, as $d\rightarrow 2^+$, $\lambda_*\rightarrow 0$ to
obtain the asymptotic freedom of the Gross-Neveu model.  Recently,
renormalizability of the large $N$ expansion of $d=3$ four-Fermi
theory has been proven using the rigorous renormalization group
technique of constructive field theory \cite{CFVS}.

In this Letter we shall assume that $\lambda_*$ exists and analyze the
resulting conformal field theory.  We find a solution of the bootstrap
equations in the three and five vertex approximation.  In particular,
we show that these give an easy way to find the anomalous dimensions
of $\psi$ and $\bar\psi\psi$ in the large $N$ limit to order $1/N^2$.

Under rescaling of the coordinate, $x\rightarrow\rho x$, the fields
transform as
\begin{equation}
\psi ^\prime(\rho x)=\rho^{-l}\psi (x)\;,~~
\bar{\psi}^\prime (\rho x)=\rho^{-l}\bar{\psi} (x)\;,~~
\phi^\prime (\rho x)=\rho^{-b}\phi (x)\;,
\end{equation}
with $l$ and $b$ the scaling dimensions of the fermion
(anti--fermions) and the boson, respectively.  Under a special
conformal transformation parameterized with a constant vector $t_\mu$,
\begin{equation}
x_\mu~\rightarrow ~x^\prime_\mu = \frac{x_\mu + t_\mu x^2}{1+2t\cdot x
+t^2x^2}\;,
\label{conf}
\end{equation}
where $\sigma_x \equiv \sigma_t(x) =|\frac{\partial x}{\partial
x^\prime}|^{1/d}= 1+2t\cdot x+t^2x^2$, they transform as
\begin{eqnarray}
\psi (x)~&\rightarrow&~\psi ^\prime(x^\prime)
=\sigma_x^{l-1/2}(1+\hat{t}\hat{x})\psi(x)\;,
\label{psi}\\
\bar{\psi}(x)~&\rightarrow&~\bar{\psi}^\prime(x^\prime)
=\sigma_x^{l-1/2}\bar{\psi}(x)(1+\hat{x}\hat{t})\;,
\label{psib}\\
\phi (x)~&\rightarrow&~\phi ^\prime(x^\prime)
=\sigma_x^b \phi (x)\;,\label{phi}
\end{eqnarray}
where $\hat{x} = \gamma_\mu x_\mu$ ($\gamma_\mu$ are the Dirac
Matrices).

These, together with translation and Lorentz invariance, form the
global finite dimensional conformal group.  Conformal invariance
determines the form of two-point correlation functions.  In momentum
space,
\begin{equation}
 G(p) = \frac{1}{i\hat{p}}(p^2)^{l-h+1/2}\;,~~~ D(p) = \frac{1}{p^2}
(p^2)^{b-h+1}\;.
\label{gd}
\end{equation}
A convenient normalization has been chosen. (This can always be done
using a finite rescaling of $\phi$ and $\psi$.)  Moreover, conformal
invariance fixes the three point vertex up to a dimensionless constant
factor, the effective coupling $\lambda$,
\begin{equation}
\Gamma(p_1,p_2)=\lambda
\frac{N(\gamma)}{N(b)N(l-b/2)}
\int \frac{d^dk}{\pi^h}\frac{\hat k +\hat p_1}{[(k+p_1)^2]^{b/2+1/2}}
\frac{\hat k +\hat p_2}{[(k+p_2)^2]^{b/2+1/2}}
\frac{1}{[k^2]^{l-b/2}}\;,
\label{gamap}
\end{equation}
where $N(\tau) \equiv \frac{\Gamma(h-\tau)}{\Gamma(\tau)}$; and
\begin{equation}
\gamma = l + b/2 - h\;,
\label{index}
\end{equation}
which is defined as the index of the vertex. By dimensional analysis,
it is easy to check that the (anomalous) dimension of the vertex
$\Gamma$ in momentum units is $-2\gamma$.

Our task is to determine the scaling dimensions $l$ and $b$, which
describe the critical behavior of the model, and the critical coupling
constant $\lambda=\lambda_*$.  We shall use the bootstrap equations.
These are derived from the Dyson-Schwinger equations for the fermion
and boson self-energies and for the Yukawa vertex.  We present the one
for the vertex first.  Graphically, the bootstrap equation of the
Yukawa vertex is
\cite{Mig71}\cite{parisi}

\unitlength=1.0mm
\linethickness{0.4pt}
\begin{picture}(140.00,51.00)
\thicklines
\put(2.00,17.00){\makebox(0,0)[cc]{$\Gamma~~~ =$}}
\put(16.00,17.00){\circle*{2.00}}
\put(16.00,17.0){\line(-1,-2){4.00}}
\put(16.00,17.0){\line(1,-2){4.00}}
\multiput(16.0,16.0)(0.00,2.00){5}{\line(0,3){1.00}}
\put(27.00,17.00){\makebox(0,0)[cc]{$=$}}
\put(50.00,30.00){\circle*{2.00}}
\put(40.50,11.00){\circle*{2.00}}
\put(59.50,11.00){\circle*{2.00}}
\multiput(50.0,29.0)(0.00,2.00){5}{\line(0,3){1.00}}
\put(50.00,30.0){\line(-1,-2){13.00}}
\put(50.00,30.0){\line(1,-2){13.00}}
\multiput(40.50,11.0)(2.00,0.00){10}{\line(3,0){1.00}}
\put(42.00,22.00){\makebox(0,0)[cc]{$k$}}
\put(42.00,5.00){\makebox(0,0)[cc]{$p_1$}}
\put(58.00,5.00){\makebox(0,0)[cc]{$p_2$}}
\put(73.00,17.00){\makebox(0,0)[cc]{$+$}}
\put(100.00,30.00){\circle*{2.00}}
\multiput(100.0,29.0)(0.00,2.00){5}{\line(0,3){1.00}}
\put(93.50,16.50){\circle*{2.00}}
\put(106.50,16.50){\circle*{2.00}}
\multiput(92.50,16.50)(2.00,-1.00){9}{\line(3,0){1.00}}
\multiput(106.50,16.50)(-2.00,-1.00){9}{\line(3,0){1.00}}
\put(89.50,8.50){\circle*{2.00}}
\put(110.50,8.50){\circle*{2.00}}
\put(100.00,30.0){\line(-1,-2){14.00}}
\put(100.00,30.0){\line(1,-2){14.00}}
\put(128.00,17.00){\makebox(0,0)[cc]{$+ \ldots$}}
\put(88.00,13.00){\makebox(0,0)[cc]{$k$}}
\put(112.00,13.00){\makebox(0,0)[cc]{$q$}}
\put(91.,4.0){\makebox(0,0)[cc]{$p_1$}}
\put(109.0,4.0){\makebox(0,0)[cc]{$p_2$}}
\end{picture}
%
%
\begin{description}
\item[Fig.~1]
\ \ \ \ The bootstrap equation for the Yukawa vertex. The dark spots
represent the conformal Yukawa vertex, the solid and dashed lines
represent the conformal fermion and boson propagators, respectively.
\end{description}
\vspace{0.5cm}

It is worth noting that the contribution of the bare vertex which
would normally appear in the right-hand-side of this equation is
absent.  This is a result of the infinite renormalization (and
therefore zero of the multiplicative renormalization constant) which
is necessary to make the theory finite.  Moreover, it is required by
the conformal ans\"atz, since the bare vertex does not have the
conformally invariant form given in (\ref{gamap}) \cite{Mig71}.  The
Dyson--Schwinger equation can be rewritten in a neat form by factoring
the momentum dependence from both sides.  To see this, let the
momentum carried by the external boson field be zero.  Then the
l.h.s. is readily calculated.  Performing the integration in
(\ref{gamap}) with $p_1=p_2=p$, we obtain
\begin{equation}
\Gamma(p,p) = \lambda\frac{1}{[p^2]^\gamma}\;.
\label{gamp}
\end{equation}
The r.h.s. of the bootstrap vertex equation is an infinite
series. By dimensional analysis, each term is proportional to
$\frac{1}{[p^2]^\gamma}$.  More precisely, the n-th term is
\begin{equation}
\lambda^{(2n+1)}f_{(2n+1)}(l,l,b)\frac{1}{[p^2]^\gamma}\;,
\end{equation}
where the function $f_{(2n+1)}(l,l,b)$ is given by the diagrams with
$2n+1$ conformal vertices. The arguments $l$, $l$, and $b$ refer to
the scaling dimensions of the external fermion, anti-fermion, and
boson, respectively.  The bootstrap equation for the vertex is
\begin{eqnarray}
1 &=& \lambda_*^2f(l,l,b;\lambda_*)\;,\label{ds1}\\ f(l,l,b;\lambda_*)
&=& \sum^{\infty}_{n=1}\lambda_*^{2(n-1)}
f_{(2n+1)}(l,l,b)\;.\label{f}
\end{eqnarray}
We call $f(l,l,b;\lambda_*)$ the vertex function.

There are two similar Dyson-Schwinger equations for the fermion and
boson self-energies, respectively,
\vskip 0.3truein

\unitlength=1.00mm
\linethickness{0.4pt}
\thicklines
\begin{picture}(110.00,32.00)
%
%
\put(40.00, 12.00){\makebox(0,0)[l]{$\Sigma~~=$}}
\put(56.00,12.00){\line(1,0){22.00}}
\multiput(60.100,13.250)(1.00,2.00){4}{\line(3,0){1.00}}
\multiput(72.900,13.250)(-1.00,2.00){4}{\line(3,0){1.00}}
\put(73.70,12.00){\circle*{2.00}}
\multiput(65.50,20.90)(2.00,0.00){2}{\line(3,0){1.00}}
\put(110.00, 12.00){\makebox(0,0)[l]{$(a)$}}
\end{picture}

\unitlength=1.00mm
\linethickness{0.4pt}
\thicklines
\begin{picture}(110.00,28.00)
%
%
\put(40.00, 18.00){\makebox(0,0)[l]{$\Pi~~=$}}
\multiput(59.0,18)(-2.00,0.00){3}{\line(3,0){1.00}}
\put(67.0,18.00){\circle{20.00}}
\put(74.00,18.00){\circle*{2.00}}
\multiput(74.50,18.0)(2.00,0.00){3}{\line(3,0){1.00}}
\put(110.00, 18.00){\makebox(0,0)[l]{$(b)$}}
\end{picture}
\begin{description}
\item[Fig.~2]
\ \ \ \ Dyson-Schwinger equations for the fermion (a) and boson
(b) self-energies. The vertex without a dark spot stands for the bare
Yukawa vertex, associated with the bare coupling $\lambda_0$.
\end{description}
\vspace{0.5cm}

As one might expect for primitively divergent vertices in a scale
invariant theory, these are ill-defined.  The l.h.s. is
finite.  The r.h.s. is a diverging integral times a
multiplicative renormalization constant and is the indeterminate
$0\cdot\infty$.  A way to resolve this ambiguity was suggested by
Parisi \cite{parisi}.  It involves taking derivatives of each side by
external momenta and by the scale dimensions of the operators.  The
result is the bootstrap equations for the vertex function. In the
present model, we obtain \cite{first}
\begin{eqnarray}
1&=&- \frac{\lambda^4_*}{(4\pi)^h\Gamma(h)}
\frac{N(\gamma)\tilde{N}(d-l)\tilde{N}(l)\tilde{N}^2(b/2)}
{N^2(b)N(l-b/2)}
\left.\frac{\partial f(l^{\prime},l,b;\lambda_*)}{\partial l^\prime /2}
\right|_{l^\prime=l},
\label{ds2}\\
1&=&N\,\hbox{Tr}\,{\bf 1}\frac{\lambda^4_*}{(4\pi)^h\Gamma(h)}
\frac{N(\gamma)\tilde{N}^2(b/2)N(d-b)}{N(b)N(l-b/2)}
\left.\frac{\partial f(l,l,b^\prime;\lambda_*)}{\partial b^\prime/2}
\right|_{b^\prime=b}\;,
\label{ds3}
\end{eqnarray}
where $N(\tau)$ is defined after (\ref{gamap}) and $\tilde{N}(\tau)
\equiv\frac{\Gamma(h-\tau+1/2)}{\Gamma(\tau+1/2)}$.
Also, $N$ is the number of fermion species and $\rm Tr {\bf 1}$ is the
dimension of the Dirac Matrices.

It is remarkable that, in all three equations (\ref{ds2}), (\ref{ds3})
and (\ref{ds1}), the only unknown is the vertex function
$f(l,l,b;\lambda_*)$, which contains all the necessary information to
determine the three parameters.  In particular, there is no need to
calculate any Feynman diagrams for the self-energies.

Unfortunately, calculating the vertex function $f(l,l,b;\lambda_*)$
is a difficult problem, as it involves an infinte number of Feynman
diagrams.  On the other hand (\ref{f}) is a power series in the
coupling constant and if the critical coupling is small, the first few
terms are expected to give good approximate results.
One limit in which $\lambda_*$ is small is the limit of large number of
fermion species, $N$.  There, $\lambda^2_*\sim1/N+\cdots$
and, as we shall demonstrate below,
further approximation to the bootstrap equations can be solved
analytically to order $1/N^2$.  Quantities which are also small in the
large $N$ limit are the anomalous dimensions defined by
\begin{eqnarray}
\gamma_\psi &=& l-h+1/2\;,\\
\gamma_\phi &=& b-1\;;
\end{eqnarray}
which are both of order $1/N$.  In terms of them, the index of the
vertex is
\begin{equation}
\gamma = l+b/2-h=\gamma_\psi+\gamma_\phi/2\;.
\end{equation}

The integral on the r.h.s. of the vertex (\ref{gamap}) is related to
the Appel function $F_4$ \cite{FGGP74}.  It takes a simple form, when
$p_1^2\ll p_2^2=p^2$. In this case,
\begin{equation}
\Gamma(p_1=0,p_2=p)
=\frac{\lambda}{[p^2]^\gamma}
\frac{\tilde{N}(b/2)\tilde{N}(l)}{N(b)N(l-b/2)}\;.
\label{gama2}
\end{equation}
It is easy to check that, at the lowest order in $\gamma$,
(\ref{gama2}) coincides with (\ref{gamp}).

To solve the set of conformal bootstrap equations, we start with
calculating the three-vertex correction to the vertex function.
Using the conformal propagators and vertex, and
setting the momentum carried by the external boson zero ({\it i.e.}
$p_1 = p_2$), the
three-vertex diagram in Fig.~1 is
\begin{eqnarray}
& &
-\lambda^3 \int \frac{d^dk}{(2\pi)^d}
\Gamma(p,k)
\frac{\hat{k}}{[k^2]^{h-l+1/2}}
\Gamma(k,k)
\frac{\hat{k}}{[k^2]^{h-l+1/2} }
\Gamma(k,p)
\frac{1}{[(k-p)^2]^{h-b}}\nonumber\\
&=& -\frac{\lambda^3}{(p^2)^\gamma}\frac{1}{(4\pi)^h}
[\frac{\tilde{N}(b/2)\tilde{N}(l)}{N(b)N(l-b/2)}]^2
\frac{N(\gamma+b)}{N(b)N(\gamma)}\;.
\label{f_32}
\end{eqnarray}
Above, we have used
(\ref{gama2}) for $\Gamma(p,k)$, as the
integration over the region of large internal momentum dominates.
Then we have
\begin{equation}
f_3(l,l,b) =  -\frac{1}{(4\pi)^h\Gamma(h)}\frac{1}{\gamma}
[1 - \frac{\gamma}{h-1} + ...]\;.
\label{f3}
\end{equation}

Substituting the first term of (\ref{f3}) into the set of conformal
bootstrap equations (\ref{ds1}), (\ref{ds2}), and (\ref{ds3}), we
obtain the anomalous dimenions of the fermion field $\psi$ and
composite operator $\phi = \bar\psi\psi$ at the leading order $1/N$
\begin{eqnarray}
\gamma_\psi^{(1)} &=& -\frac{1}{N{\rm Tr}{\bf 1}}
\frac{\Gamma(2h-1)\sin(\pi h)}{\pi\Gamma(h-1)\Gamma(h+1)}
\;,
\label{gpsi1}\\
\gamma_\phi^{(1)}&=&  -\frac{2(2h-1)}{h-1}\gamma_\psi^{(1)}
\;,
\label{gphi1}
\end{eqnarray}
which reproduce the known results,
with
\begin{eqnarray}
(\lambda^2_*)^{(1)}&=&
-\frac{1}{N{\rm Tr}{\bf 1}}
\frac{(4\pi)^h\Gamma(2h-1)\sin(\pi h)}{\pi\Gamma(h)}
\;,
\\
\gamma^{(1)}&=&\frac{1}{N{\rm Tr}{\bf 1}}
\frac{\Gamma(2h-1)\sin(\pi h)}{\pi\Gamma^2(h)}
\;.
\end{eqnarray}

Moreover, to the next to leading order, as we shall see below, the
vertex function $f$ takes a same form as that in (\ref{f3}).  This
implies to the order
\begin{equation}
\left.\frac{\partial f(l^{\prime},l,b;\lambda_*)}{\partial l^\prime /2}
\right|_{l^\prime=l}=
\left.\frac{\partial f(l,l,b^\prime;\lambda_*)}{\partial b^\prime /2}
\right|_{b^\prime=b}\;.
\end{equation}
Then, with no need of further details of the vertex function
$f(l,l,b)$, by using the boostrap equations (\ref{ds2}) and
(\ref{ds3}), one can determine the fermion anomalous dimenion to order
$1/N^2$. Dividing one of them by the other, we have
\begin{equation}
1 = -\frac{1}{NTr{\bf 1}}
\frac{\tilde{N}(l)\tilde{N}(d-l)}{N(b)N(d-b)}\;.
\end{equation}
Expand the r.h.s. of the above equation over $\gamma_\psi$ and
$\gamma_\phi$ (both $\sim 1/N$+...), we obtain
\begin{eqnarray}
\gamma_\psi&=&\gamma^{(1)}_\psi[1+\frac{\gamma_\psi^{(1)}}{h}
-\gamma_\phi^{(1)}(\frac{1}{h-1} + \psi(2h-1)-\psi(1) + \pi ctg(h\pi))
+ ...]
\;,
\label{gamapsi}
\end{eqnarray}
where $\psi(x) = \Gamma^\prime (x)/\Gamma(x)$, and $\gamma_\psi^{(1)}$
and $\gamma_\phi^{(1)}$ are given in (\ref{gpsi1}) and (\ref{gphi1}).

To obtain the anomalous dimension of $\phi = \bar\psi\psi$ to order
$1/N^2$, one needs the vertex function $f$ to the same order.
Besides the second term in the three-vertex correction (\ref{f3}),
it involves the leading order of the five-vertex correction, and
also the leading order of the seven-vertex diagrams with a fermion loop.

Notice that only one five-vertex diagram, depicted in
Fig.~1, contributes to $f_5(l,l,b)$.
(Another diagram which also contains five
conformal vertices has a subdiagram with a fermion loop attached to
three boson legs and is zero by parity symmetry).  The Feynman
integral (setting $p_1 = p_2$) is
\begin{eqnarray}
& & \int \frac{d^dq}{(2\pi)^d}
         \int       \frac{d^dk}{(2\pi)^d}
\Gamma(p,k)
G(k)
\Gamma(k,k+q-p)
G(k+q-p)
\Gamma(k+q-p,k+q-p)
\nonumber\\
& &
{}~~~~~~~~~~~~~~~~G(k+q-p)
\Gamma(k+q-p,q)
G(q)
\Gamma(q,p)
D(k-p)
D(k-q)\;.
\label{f_5}
\end{eqnarray}
Performing the integral, to the leading logarithmic term, we obtain
\begin{equation}
f_5(l,l,b) =  -\frac{1}{(4\pi)^d\Gamma^2(h)(h-1)}
\frac{1}{\gamma} + ...\;.
\label{f5}
\end{equation}

To order $1/N^2$, the seven-vertex diagrams with a fermion loop
attached to four boson lines contribute, as a fermion loop carries
a factor $N$. This sort of seven-vertex diagrams are given in Fig.~3.

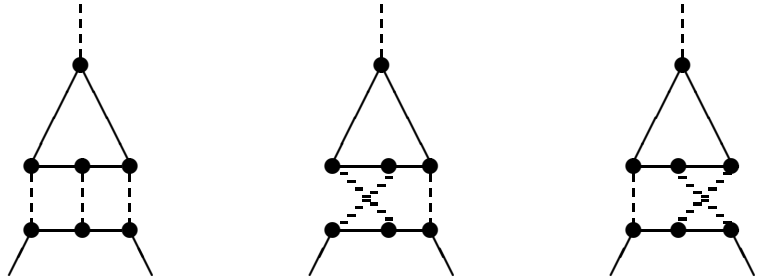
\unitlength=1.0mm
\linethickness{0.4pt}
\begin{picture}(140.00,43.00)
\thicklines
\put(30.00,30.00){\circle*{2.00}}
\multiput(30.0,29.0)(0.00,2.00){5}{\line(0,3){1.00}}
\put(23.50,16.50){\circle*{2.00}}
\put(36.50,16.50){\circle*{2.00}}
\put(23.50,16.50){\line(1,0){12.00}}
\put(23.50,8.0){\circle*{2.00}}
\put(36.50,8.0){\circle*{2.00}}
\put(30.25,16.50){\circle*{2.00}}
\put(30.25,8.0){\circle*{2.00}}
\put(23.50,8.0){\line(1,0){12.00}}
\multiput(23.50,8.50)(0.00,2.00){5}{\line(0,3){1.00}}
\multiput(30.250,8.50)(0.00,2.00){5}{\line(0,3){1.00}}
\multiput(36.50,8.50)(0.00,2.00){5}{\line(0,3){1.00}}
\put(30.00,30.0){\line(-1,-2){7.00}}
\put(30.00,30.0){\line(1,-2){7.00}}
\put(23.50,8.0){\line(-1,-2){3.00}}
\put(36.50,8.0){\line(1,-2){3.00}}
\put(70.00,30.00){\circle*{2.00}}
\multiput(70.0,29.0)(0.00,2.00){5}{\line(0,3){1.00}}
\put(63.50,16.50){\circle*{2.00}}
\put(76.50,16.50){\circle*{2.00}}
\put(63.50,16.50){\line(1,0){12.00}}
\multiput(63.50,16.50)(1.00,-1.00){8}{\line(3,0){1.00}}
\multiput(63.50,8.0)(1.00,1.00){8}{\line(3,0){1.00}}
\put(63.50,8.0){\circle*{2.00}}
\put(76.50,8.0){\circle*{2.00}}
\put(71.0,16.50){\circle*{2.00}}
\put(71.0,8.0){\circle*{2.00}}
\put(63.50,8.0){\line(1,0){12.00}}
\multiput(76.50,8.50)(0.00,2.00){5}{\line(0,3){1.00}}
\put(70.00,30.0){\line(-1,-2){7.00}}
\put(70.00,30.0){\line(1,-2){7.00}}
\put(63.50,8.0){\line(-1,-2){3.00}}
\put(76.50,8.0){\line(1,-2){3.00}}
\put(110.00,30.00){\circle*{2.00}}
\multiput(110.0,29.0)(0.00,2.00){5}{\line(0,3){1.00}}
\put(103.50,16.50){\circle*{2.00}}
\put(116.50,16.50){\circle*{2.00}}
\put(103.50,16.50){\line(1,0){12.00}}
\multiput(116.50,16.50)(-1.00,-1.00){8}{\line(3,0){1.00}}
\multiput(116.50,8.0)(-1.00,1.00){8}{\line(3,0){1.00}}
\put(103.50,8.0){\circle*{2.00}}
\put(116.50,8.0){\circle*{2.00}}
\put(109.50,16.50){\circle*{2.00}}
\put(109.50,8.0){\circle*{2.00}}
\put(103.50,8.0){\line(1,0){12.00}}
\multiput(103.50,8.50)(0.00,2.00){5}{\line(0,3){1.00}}
\put(110.00,30.0){\line(-1,-2){7.00}}
\put(110.00,30.0){\line(1,-2){7.00}}
\put(103.50,8.0){\line(-1,-2){3.00}}
\put(116.50,8.0){\line(1,-2){3.00}}
\end{picture}
%
%
\begin{description}
\item[Fig.~3]
\ \ \ \ The diagrams of seven-vertex correction with a fermion loop.
Each fermion loop carries a factor $N$.
\end{description}
\vspace{0.5cm}

For $f_7$ in the second order, to the leading logarithmic term again,
we have
\begin{eqnarray}
& &  \frac{N{\rm Tr}{\bf 1}~\Gamma(2-h)}
                   {(4\pi)^{3h}(h-1)^2\Gamma(2h-2)}
\frac{1}{\gamma}[1-\frac{3\Gamma^3(h/3)\Gamma(h)(2h/3-1)}
{8h\Gamma^3(2h/3)}] + ...\;,\label{f71}\\
& &  \frac{N{\rm Tr}{\bf 1}~\Gamma(2-h)}
                   {(4\pi)^{3h}(h-1)^2\Gamma(2h-2)}
\frac{1}{\gamma}[1+\frac{1}{h-1} -\frac{\Gamma^3(h/3)\Gamma(h)}
{2\Gamma^3(2h/3)}] + ...\;,
\label{f72}
\end{eqnarray}
for the first and the next two diagrams in Fig.~$3$, respectively.
The vertex function is
\begin{equation}
f(l,l,b, \lambda) = f_3(l,l,b)
+ \lambda^2_*f_5(l,l,b)+\lambda^4_*f_7(l,l,b)+ ...\;,
\label{fff}
\end{equation}
where $f_3$, $f_5$, and $f_7$, to the second order, are given in
(\ref{f3}), (\ref{f5}), and (\ref{f71}) and (\ref{f72}).  This is a
systematic expansion procedure. To calculate the next order, $1/N^3$,
for instance, requires the first three terms in the expansion of
$f_3$, the first two terms of $f_5$, and the first two terms of $f_7$
(one with a factor $N$ carried by a fermion loop and the other
without), and the first term of $f_9$ with an explicit $N$ factor.
Similarly, as the multiplicative factors of the partial derivatives of
$f(l,l,b;\lambda_*)$ of equations (\ref{ds2}) and (\ref{ds3}) are
functions of $l$ and $b$, we expand these to the second order in
$\gamma_\psi$ and $\gamma_\phi$ (or $\gamma$) as well.  Substituting
these expansions in the equation (\ref{ds2}) and (\ref{ds1}), we
obtain
\begin{eqnarray}
\lambda^2_*&=& (\lambda^2_*)^{(1)}
[1
-\gamma_\phi^{(1)}(\pi ctg(\pi h)+\psi(2h-1)-\psi(1)) + ...]\;,
\;\\
\gamma& =& \gamma^{(1)}
[1 - \frac{\gamma^{(1)}}{h-1}
+\frac{{\lambda^2_*}^{(1)}}{(4\pi)^h\Gamma(h)(h-1)}
-\gamma_\phi^{(1)}(\pi ctg(\pi h)+\psi(2h-1)-\psi(1))
\nonumber\\
& &
- \frac{({\lambda^2_*}^{(1)})^2N{\rm Tr}{\bf 1}~\Gamma(2-h)\Gamma(h)}
{(4\pi)^{2h}\Gamma(2h-2)(h-1)^2}
(\frac{2h-1}{h-1} - \frac{\Gamma^3(h/3)\Gamma(h)}
{2\Gamma^3(2h/3)}(1+\frac{1}{2}(1 - \frac{3}{2h}))
+ ...]\;. \nonumber\\
& &
\label{gama}
\end{eqnarray}
Then finally,
\begin{equation}
\gamma_\phi = 2(\gamma - \gamma_\psi)\;,
\end{equation}
where $\gamma$ and $\gamma_\psi$ are given in (\ref{gama}) and (\ref{gamapsi}).

In particular, when $d=2h=3$ and ${\rm Tr}{\bf 1} = 2$, we have
\begin{eqnarray}
\gamma_\psi &=&\frac{2}{3\pi^2N}[1 + \frac{148}{9\pi^2N} + ...]\;,\\
\gamma_\phi &=&-\frac{16}{3\pi^2N}[1 + \frac{76+27\pi^2}{36\pi^2N} + ...]\;.
\end{eqnarray}

The $\psi$ anomalous dimension ($\gamma_\psi$) has been computed to
this order by conventional diagrammatic methods in \cite{G}.  That
calculation involves much more labor than the present one.  Also, the
present calculation gives the anomalous dimension of the composite
operator $\phi=\bar\psi\psi$ which has not been computed elsewhere.
Note that the result for $\gamma_\psi$ is not the same as in \cite{G}.
(The numerator of the second term quoted in \cite{G} has 122 instead
of 148.)  We presently do not understand the source of this
discrepancy.

Here, we see the power of the bootstrap method in performing
approximate calculations, such as the large $N$ expansion to higher
orders.  It would be interesting to apply this approach to other field
theories, such as gauge theory or theories with vector-like vertices.
It would also be interesting to search for other, non--perturbative
solutions in the present four--fermion, or in $\lambda\phi^4$ theory
\cite{Mak79}.

\vspace{0.8cm}

Note added: After this work was submitted for publication, J A Gracey
informed the authors that the number 122 was a misprint in \cite{G},
his result would be 112, instead.  In a recent publication \cite{G1}
he also calculated the exponent $\gamma_\phi$ to order $1/N^2$.

\end{document}